 \documentclass[preprint,showpacs,preprintnumbers,superscriptaddress,amsmath,amssymb]{revtex4}

 \usepackage[dvips]{graphicx}
 \usepackage{mathrsfs}
 \usepackage{amssymb}
 \usepackage{dcolumn}
 \usepackage{bm}
 \usepackage{subfigure}
 \usepackage{floatflt}
 \usepackage{float}
 \usepackage{rotating}
 \usepackage{multirow}
 \usepackage{CJK}

\begin{document}
 \preprint{00-000}

 \title{In-medium nucleon-nucleon elastic cross sections determined from nucleon-induced reaction cross sections data}

 \author{Li Ou}
 \email{liou@gxnu.edu.cn}
 \affiliation{College of Physics and Technology, Guangxi Normal University, Guilin, 541004, P. R. China}
 \affiliation{Guangxi Key Laboratory of Nuclear Physics and Technology, Guangxi Normal University, Guilin 541004, China}

 \author{Xueying He}
 \affiliation{College of Physics and Technology, Guangxi Normal University, Guilin, 541004, P. R. China}
 \affiliation{Guangxi Key Laboratory of Nuclear Physics and Technology, Guangxi Normal University, Guilin 541004, China}

 \date{\today}

 \begin{abstract}

 Within the framework of the improved quantum molecular dynamics model,
 the medium modifications on the free nucleon-nucleon elastic cross sections
 are investigated.
 By using various in-medium nucleon-nucleon elastic cross sections in the model,
 the nucleon-induced reactions on various targets are simulated,
 and the excitation functions of reaction cross sections in the energy range from 25 MeV to 1 GeV are calculated.
 By comparing the calculations with the experimental data, an isospin, density, and momentum-dependence
 medium correction factor on free nucleon-nucleon elastic cross sections is determined.

 \end{abstract}

 \maketitle

 \section{Introduction}

 With more and more upcoming experimental data of rare isotope heavy ion collision at intermediate energies,
 people have opportunity to explore the properties of nuclear matter with high density and large isospin asymmetry
 by the tools of transport model.
 Both the mean field and the in-medium two-body scattering cross sections,
 coming from the same nucleon-nucleon (NN) interaction \cite{Danielewicz1984a,Danielewicz1984b,ChouKC1985,HanYL1994},
 should be treated self-consistently in the transport model.
 But it is very difficult to solve the dynamical equation and the G-matrix simultaneously,
 the most of transport models deal with the mean field and collision separately.
 So the accurate in-medium nucleon-nucleon cross sections (NNCS) are required by transport model.
 On the other hand, the in-medium NNCS can provide information
 which is very important to study the structure of nuclei especially rare isotopes.
 Also the in-medium NNCS are of interest for their own sake,
 as they are underly related to the viscosity, the mean free path of nucleon in nuclear matter,
 and other nuclear transport coefficients \cite{Cugnon1987,Kohler1991,Danielewicz2003}.

 Unlike the NNCS in the free space which can be directly measured by experiment,
 the in-medium NNCS can only be calculated by theory,
 such as the Brueckner-Hartree-Fock theory, the Dirac-Brueckner-Hartree-Fock theory,
 the Green function approach, and the relativistic mean field model.
 The in-medium NNCS depend on several quantities, such as density and isospin-asymmetry of nuclear matter,
 total momentum of the nucleon pair in the nuclear matter rest frame, the relative momentum of the nucleon pair,
 even the temperature.
 There have been many studies demonstrate that the NNCS in medium should be suppressed compared to free NNCS
 \cite{ChouKC1985,MaoGJ1994,MaoGJ1996,LiQF2000,LiQF2004,Haar1987,Haar1987R,LiGQ1993,LiGQ1994,Fuchs2001,Kohno1998,Alm1994,Sammarruca2014}.
 However, the modification factors given from these works differ significantly,
 due to the different treatment on the dependence with different levels of approximations.

 The other way to get the information of the in-medium NNCS is
 to compare the simulation results from transport theory with experimental data of nuclear reactions.
 As the basic ingredients of transport model, some kind of simple parametrizations of the in-medium NNCS are widely used.
 Such as the parameterizations given by G. Li \cite{LiGQ1993,LiGQ1994} and Q. Li \cite{LiQF2006,LiQF2011,WangYJ2014},
 the phenomenological in-medium NNCS scaled by effective mass \cite{Negele1981,Pandharipande1991,Persram2002,LiBA2005}
 \begin{eqnarray}
 \sigma^*_{\rm NN}=\left[ \frac{\mu^*(\rho, p)}{\mu(p)} \right]^2\sigma^{\rm free}_{\rm NN},
 \end{eqnarray}
 where $\mu$ and $\mu^*$ are the reduced masses of colliding nucleon pairs in the vacuum and medium, respectively,
 and the empirical relation
 \begin{eqnarray}
 \sigma^*_{\rm NN}=( 1+\eta \rho/\rho_0 )\sigma^{\rm free}_{\rm NN},
 \end{eqnarray}
 where $\eta$ is an adjustable parameter.
 Consistent with theory results, the most of experimental evidences, such as balance energy \cite{Westfall1993,ZhouHB1993,Klakow1993},
 stopping power \cite{ZhangYX2006}, collective flow \cite{Persram2002,ZhangYX2006,ZhangYX2007,WangYJ2014}, etc,
 support a reduced in-medium nucleon-nucleon elastic cross sections (NNECS).
 But the reduced factor has not been determined very definitely.
 Since both the mean field and the two body collisions are convoluted,
 model dependent treatment of the nuclear potential as well as the collisions may yield considerable discrepancies
 in the model outputs even for the simplest box calculations.
 Compared to the heavy ion collision, some kind of direct reactions involve less degrees of freedom
 in the reaction process may reduce the model dependence.
 So the nucleon-induced reaction is more suitable to study the in-medium NNCS,
 not only because of its simple mechanism but also
 because it reflects just the in-medium NNCS around the incident beam energy and below the saturation density
 \cite{Tanaka1996,Tripathi1998,Tripathi1999,Tripathi2001,OuL2008JPG}.

 In this work, with the improved quantum molecular dynamics (ImQMD) model,
 we try to extract the in-medium NNECS from the experimental data of the nucleon-induced
 reaction cross sections (RCSs).
 The paper is organized as follows.
 In Sec. II, we briefly introduce the model we adopted.
 In Sec. III, we introduce the form of medium modification on NNCS,
 and the way to determine the modification factor.
 Finally a brief summary is given in Sec. IV.

 \section{Model}

 In the frame of ImQMD05 model \cite{ZhangYX2006,OuL2008PRC}, each nucleon is described as a Gaussian wave-packet.
 In the mean filed, the motions of centers of the wave-packets follow the evolution of Hamilton canonical equation.
 The potential energy $U$ including local nuclear potential energy and Coulomb energy, reads
  \begin{eqnarray}\label{Eqnarry3}
 U=U_{\rm{loc}}+U_{\rm{Coul}}.
 \end{eqnarray}
 The local part is obtained from the Skyrme interaction, $U_{\rm{loc}}=\int V_{\rm{loc}}[\rho(\bm{r})] d \bm{r}$.
 $V_{\rm{loc}}$ is the Skyrme potential energy density functional with only the spin-orbit term omitted, which reads
\begin{equation}
\label{Eqnarry4}
V_{\rm{loc}}=\frac{\alpha}{2}\frac{\rho ^{2}}{\rho _{0}}+\frac{\beta }{\eta +1}%
 \frac{\rho ^{\eta +1}}{\rho _{0}^{\eta }}+\frac{g_{\rm{sur}}}{2\rho _{0}}\left(\nabla \rho \right)^{2}
 +\frac{g_{\rm{sur,iso}}}{\rho_{0}}[\nabla(\rho_{\rm{n}}-\rho_{\rm{p}})]^{2}
 +(A\rho^{2}+B\rho^{\eta+1}+C\rho^{8/3})\delta^{2}+g_{\rho\tau}\frac{\rho^{8/3}}{\rho_{0}^{5/3}},
\end{equation}
 The subscript symbols ``$\rm n$'' and ``$\rm p$'' indicate the neutron, and
 proton respectively. And ``$\delta$'' is the isospin asymmetry $\delta = (\rho_{n}-\rho_{p})/(\rho_{n}+\rho_{p})$.
 The coefficients in Eq. \eqref{Eqnarry4} can be transcribed from the parameters of the standard Skyrme interactions,
 one can refer to \cite{ZhangYX2006,OuL2008PRC} for details.

 In this work, the isospin-dependent NNCS provided by Cugnon {\it et al}.~\cite{Cugnon1996} is adopted
 as free NNCS, in the collision term.
 The reaction cross sections depend directly on the collision term,
 so the isospin-dependent Pauli blocking effect should be carefully considered.
 The final states of each pair of nucleons undergoing collision must satisfy the uncertainty principle
 \begin{eqnarray}\label{Eqnarry9}
 \frac{4\pi}{3} r_{ij}^{3} \cdot \frac{4\pi}{3} p_{ij}^{3} \geq
 \frac{h^{3}}{8}.
 \end{eqnarray}
 The $r_{ij}$ and $p_{ij}$ are the distances between two nucleons in the coordinate and momentum space.
  The probabilities for a state being occupied can be calculated by
 \begin{eqnarray}\label{Eqnarry10}
 P_{i}=\sum_{k,k\neq i}^{A} \frac{1}{(\pi \hbar)^{3}}
 \exp\left[-\frac{(\bm{r}_{i}-\bm{r}_{k})^{2}}{2\sigma_{r}^{2}}\right]
 \exp\left[-\frac{(\bm{p}_{i}-\bm{p}_{k})^{2}}{2\sigma_{p}^{2}}\right].
 \end{eqnarray}
 For two nucleons scatted to the final state $i$ and $j$,
 the Pauli block possibility is
 \begin{eqnarray}\label{Eqnarry10}
 P_{\rm{block}}=1-(1-P_{i})(1-P_{j}).
 \end{eqnarray}

 The RCS for nucleon-induced reactions can be calculated by
 \begin{eqnarray}
 \sigma_R=\int_0^{b_{\rm{max}}} 2\pi P_{\rm{inel}}(b) b d b,
 \end{eqnarray}
 where $P_{\rm{inel}}(b)$ is the probability of the inelastic scattering events with the impact parameter $b$.
 $b_{\rm{max}}$ is the maximum impact parameter, i.e., there is no more inelastic collision when $b > b_{\rm{max}}$.
 On the experiment, $P_{\rm{inel}}(b)$ is indirectly determined by measuring the probability of the elastic scattering events, i.e,
 $P_{\rm{inel}}(b)$ is defined as $P_{\rm{inel}} = 1-P_{\rm{un}}-P_{\rm{el}}$.
 Any scattering event in which the emitted nucleon has energy close to the incident energy,
 whatever it has been collided or not,
 is defined as elastic scattering event.
  where $P_{\rm{un}}$ is the probability of the unaffected events,
 i.e., the events for the incident nucleon only passing through target keeping its momentum direction and magnitude,
 and $P_{\rm{el}}$ is the probability of the elastic events,
 i.e., the events for the incident nucleon only changing its momentum direction but keeping its momentum magnitude,
 respectively.
 The calculation of RCS in this work follow the analyzing method used in experiments.

 \section{Results and discussion}

\begin{figure}[hbtp]
  \centering
   \includegraphics[width=0.45\textwidth]{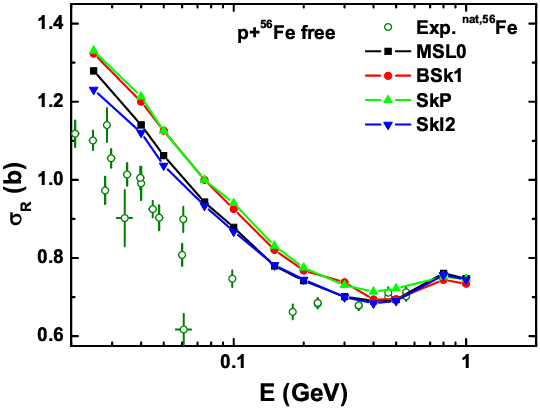}
   \caption{(Color online) Excitation functions of RCS for $p+^{56}$Fe calculated with free NNECS and
   four Skyrme parameter sets compared with experimental data. Data are taken form Ref.~\cite{Carlson1996}.}
   \label{EOS}
  \end{figure}

 As mentioned above, the mean field and NN collision affect the nuclear reaction process together.
 The effect of the mean field on RCS should be checked firstly. As a test, four Skyrme parameter sets MSL0 \cite{ChenLW2010},
 BSk1 \cite{Samyn2002}, SkP~\cite{Dobaczewski1984}, SkI2~\cite{Reinhard1995} are adopted to simulate the reactions of $p+^{56}$Fe.
 In the calculations, the free isospin dependent NNCS provided by Cugnon {\it et al}.~\cite{Cugnon1996} is adopted.
 The comparison between the calculated excitation functions of RCS and the experimental data is illustrated in Fig.~\ref{EOS}.
 One can see that, with various EOS adopted, the calculated RCSs are different.
 With energy increasing the difference becomes smaller, due to the NN collision dominates the reaction.
 For the case of energies higher than 500 MeV, the results calculated with free NNECS can reproduce the experimental data well.
 But all calculation results overestimate the experimental data below 500 MeV.
 It means that the in-medium NNECS below 500 MeV should be smaller than free NNECS
 and the in-medium NNECS above 500 MeV should be very close to free NNECS.
 To obtain correct medium correction on NNECS, the reasonable EOS tested by other method should be used in simulations.
 MSL0, one of Skyrme parameter sets which best satisfy the current understanding of
 the physics of nuclear matter over a wide range of applications~\cite{Dutra2012}, is used in the following calculations.

 According to the above analysis and the other studies on NNSC, in this work, the free NNCS is modified according to
 \begin{eqnarray}
 \sigma^{\rm{*}}_{\rm{tot}}=\sigma^{\rm{free}}_{\rm{in}}+\sigma^*_{\rm{el}}
 =\sigma^{\rm{free}}_{\rm{in}}+F(u,\delta,p)\sigma^{\rm{free}}_{\rm{el}}.
 \end{eqnarray}
 Where the $\sigma^{\rm{free}}_{\rm{el}}$ and $\sigma^{\rm{free}}_{\rm{in}}$
 are the free isospin dependent elastic and inelastic cross sections, respectively.
 The form of $F(u,\delta,p)$ is as the same as that proposed by Q. Li {\it et al}. in Refs.~\cite{LiQF2006,LiQF2011,WangYJ2014}.
 The medium correction factor $F=F_{\delta}^{p} \cdot F_{u}^{p}$ depends on
 the nuclear-reduced density $u=\rho/\rho_0$, the isospin-asymmetry
 $\delta= (\rho_{n}-\rho_{p})/(\rho_{n}+\rho_{p})$ and the momentum.
 Where
 \begin{eqnarray}
 F_u=\lambda+(1-\lambda)\exp(-u/\zeta),
 \end{eqnarray}
 \begin{eqnarray}
 F_{\delta}=1-\tau_{ij} A(u)\delta,~~~~A(u)=\frac{0.85}{1+3.25u}.
 \end{eqnarray}
 When $i=j=n,~\tau_{ij}=-1;~i=j=p,~\tau_{ij}=+1;~i\ne j,~\tau_{ij}=0$.
 The $F_{\delta}^{p}$ and $F_{u}^{p}$ factors are expressed in one formula,
 \begin{equation}\label{Eqnarry12}
 F_{\delta,u}^{p}=\left\{
 \begin{array}{lr}
 f_0, & p_{\rm{NN}}> 1~{\rm GeV}/c, \\
 \frac{F_{\delta,u}-f_0}{1+(p_{\rm{NN}}/p_0)^{\kappa}}+f_0,~~~~~~~~ & p_{\rm{NN}} \leq 1~{\rm GeV}/c,
 \end{array}
 \right.
 \end{equation}
 with $p_{\rm{NN}}$ being the relative momentum in the NN center-of-mass system.
 By varying the parameters $\lambda$, $\zeta$, $f_0$, $p_0$ and $\kappa$ one can obtain
 different medium correction on NNECS. The parameter sets used in this work
 are listed in Table \ref{table1} and \ref{table2}.
 Among these parameter sets, FU1-3 and FP1-5 are taken from Refs.~\cite{LiQF2006,LiQF2011,WangYJ2014},
 FU4 and FP6 are obtained by this work.
 \begin{table}[hbtp]
 \caption{\label{table1}
 Parameter sets used for the density-dependent correction factor $F_{u}$.}
 \begin{ruledtabular}
 \begin{tabular}{ccc}
 Set   &  $\lambda$  &  $\zeta$\\ \hline
 FU1   &   1/3 &  0.54568   \\
 FU2   &   1/4 &  0.54568   \\
 FU3   &   1/6 &  1/3   \\
 FU4   &   1/5 &  0.45   \\
 \end{tabular}
 \end{ruledtabular}
 \end{table}
 \begin{table}[hbtp]
 \caption{\label{table2}
 Parameter sets used for the momentum dependence of correction factor $F_{u,\delta}$.}
 \begin{ruledtabular}
 \begin{tabular}{cccc}
 Set   &  $f_0$  &  $p_0~[\rm{GeV/c}]$  & $\kappa$\\ \hline
 FP1   &   1  &  0.425   &  5   \\
 FP2   &   1  &  0.225   &  3   \\
 FP3   &   1  &  0.625   &  8   \\
 FP4   &   1  &  0.3  &  8   \\
 FP5   &   1  &  0.34 &  12   \\
 FP6   &   1  &  0.725   &  10   \\
 \end{tabular}
 \end{ruledtabular}
 \end{table}

\begin{figure}[hbtp]
 \centering
 \includegraphics[width=0.45\textwidth]{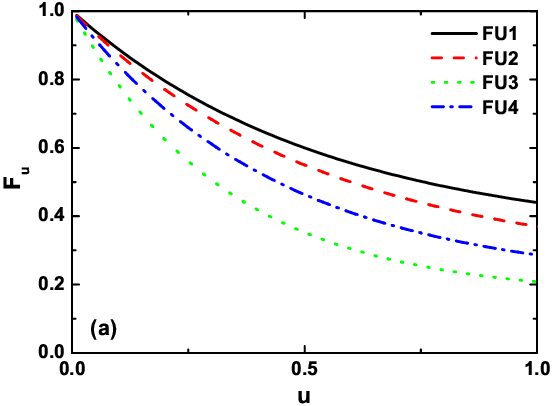}
 \includegraphics[width=0.45\textwidth]{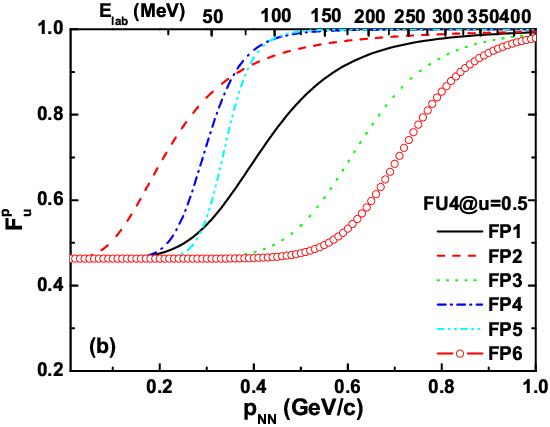}
 \caption{(Color online) (a) The density dependence of $F_u$ and
 (b) the momentum dependence of $F_u$ with FU4
 at $\rho/\rho_0=0.5$, respectively.}
 \label{Factor}
 \end{figure}

 The medium correction factor $F_u$ as function of reduced density $u$ and
 momentum dependence of $F_u$ with FU4 at $\rho/\rho_0=0.5$ are represented in Fig.~\ref{Factor}.
 From the figure, one can see the rough character of NNECS, with density increasing in-medium NNECS decrease.
 And for a certain set of density dependence, with momentum increasing the in-medium NNECS might be enhanced.

 The various in-medium NNECS obtained from combinations by parameterizations FU1, FU2, FU3 and FP1, FP2, FP3
 are tested by the excitation function of RCS for $p+^{56}$Fe.
 \begin{figure*}[hbtp]
  \centering
   \includegraphics[width=0.45\textwidth]{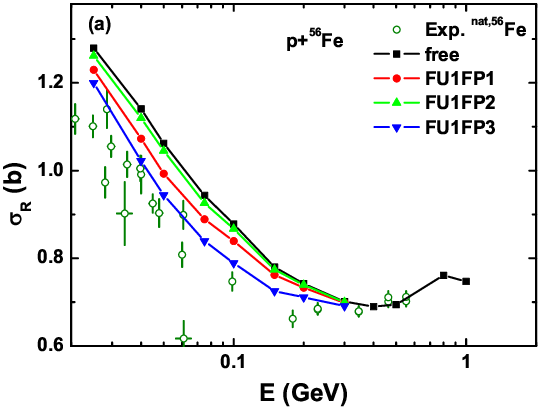}
   \includegraphics[width=0.45\textwidth]{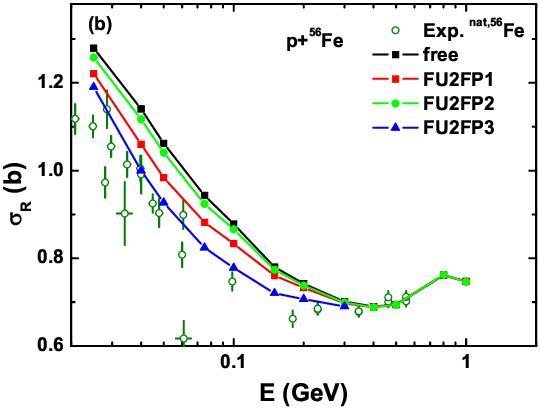}\\
   \includegraphics[width=0.45\textwidth]{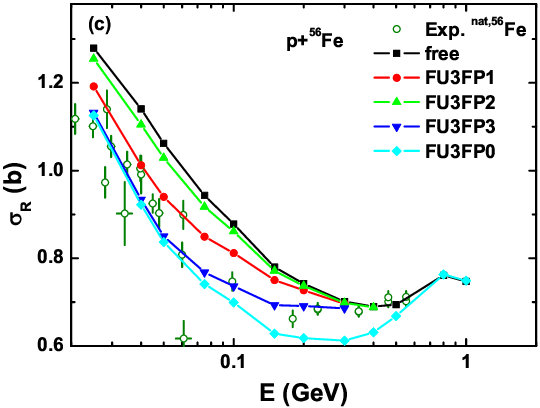}
   \includegraphics[width=0.45\textwidth]{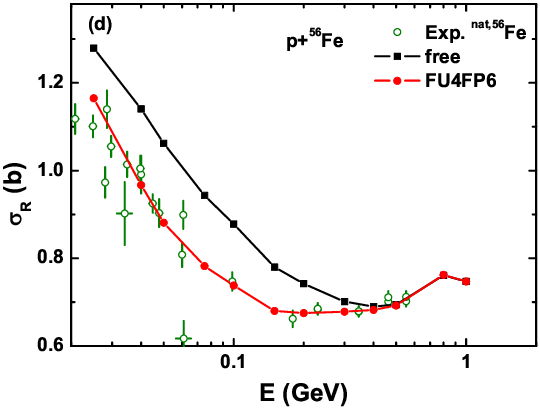}
   \caption{(Color online) Excitation functions of RCS for $p+^{56}$Fe calculated with free NNECS and
   various in-medium NNECS compared with experimental data, respectively. Experimental data are taken form Ref.~\cite{Carlson1996}.}
   \label{pFeFu}
 \end{figure*}
 By using the in-medium NNECS, the descriptions on the excitation function of RCS are great improved.
 Especially the combinations of FU2+FP3 and FU3+FP3 give the best two results except a little deviation at low energies.
 According to the momentum dependence of the in-medium NNECS $F_u^p$ shown in Fig.~\ref{Factor} (b),
 the enhancement effect of momentum correction dose not yet appear at such low energies if FP3 is adopted.
 As a test, the excitation function of RCS calculated by FU3 and without momentum correction
 is shown in subfigure (c) marked by FU3FP0.
 One can see that the momentum correction dose not obviously affects the results below 50 MeV,
 while the RCSs at high energies are underestimated if the momentum correction is absent.
 It proves our assumption and means the depressive effect provided by FU3 is a little strong while the one provided by FU2 is a little weak.
 So we propose a parameter set, namely FU4, to provides a reasonable correction effect between ones given by FU2 and FU3.
 For the density correction parameter set FU4, corresponding momentum correction parameter FP6
 is obtained by fitting the experimental data of RCSs at higher energies.

\begin{figure*}[hbtp]
  \centering
   \includegraphics[width=0.32\textwidth]{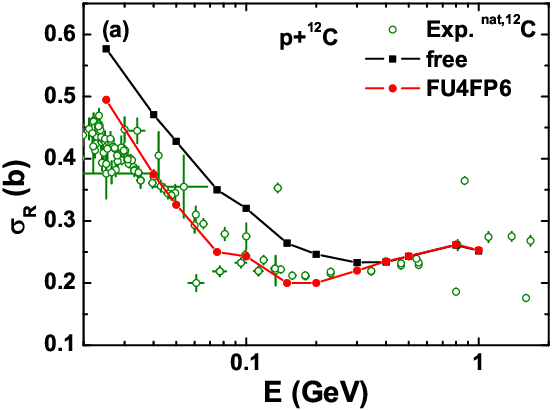}
   \includegraphics[width=0.32\textwidth]{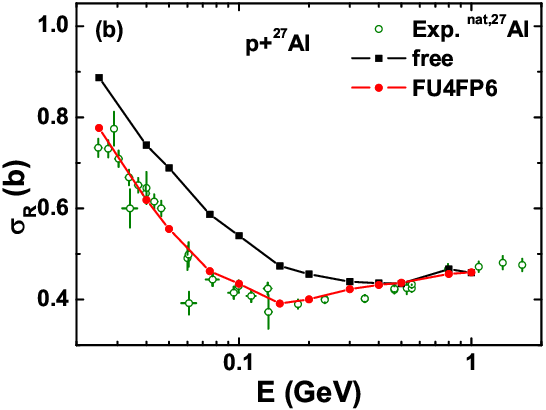}
   \includegraphics[width=0.32\textwidth]{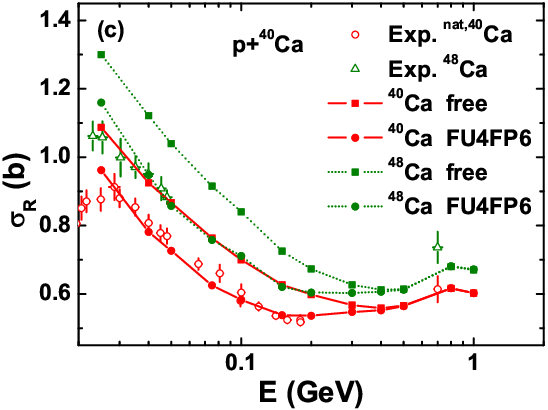}\\
   \includegraphics[width=0.32\textwidth]{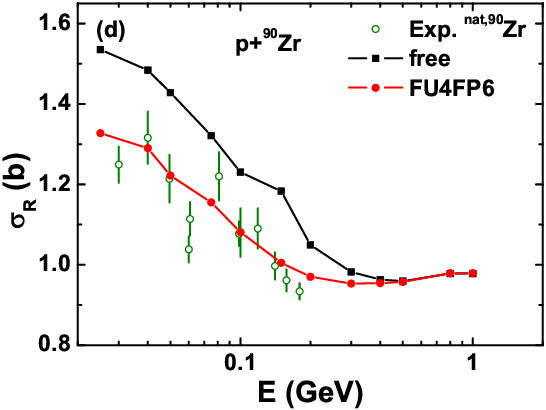}
   \includegraphics[width=0.32\textwidth]{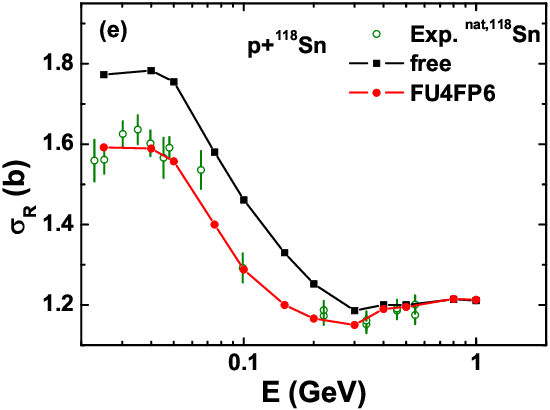}
   \includegraphics[width=0.32\textwidth]{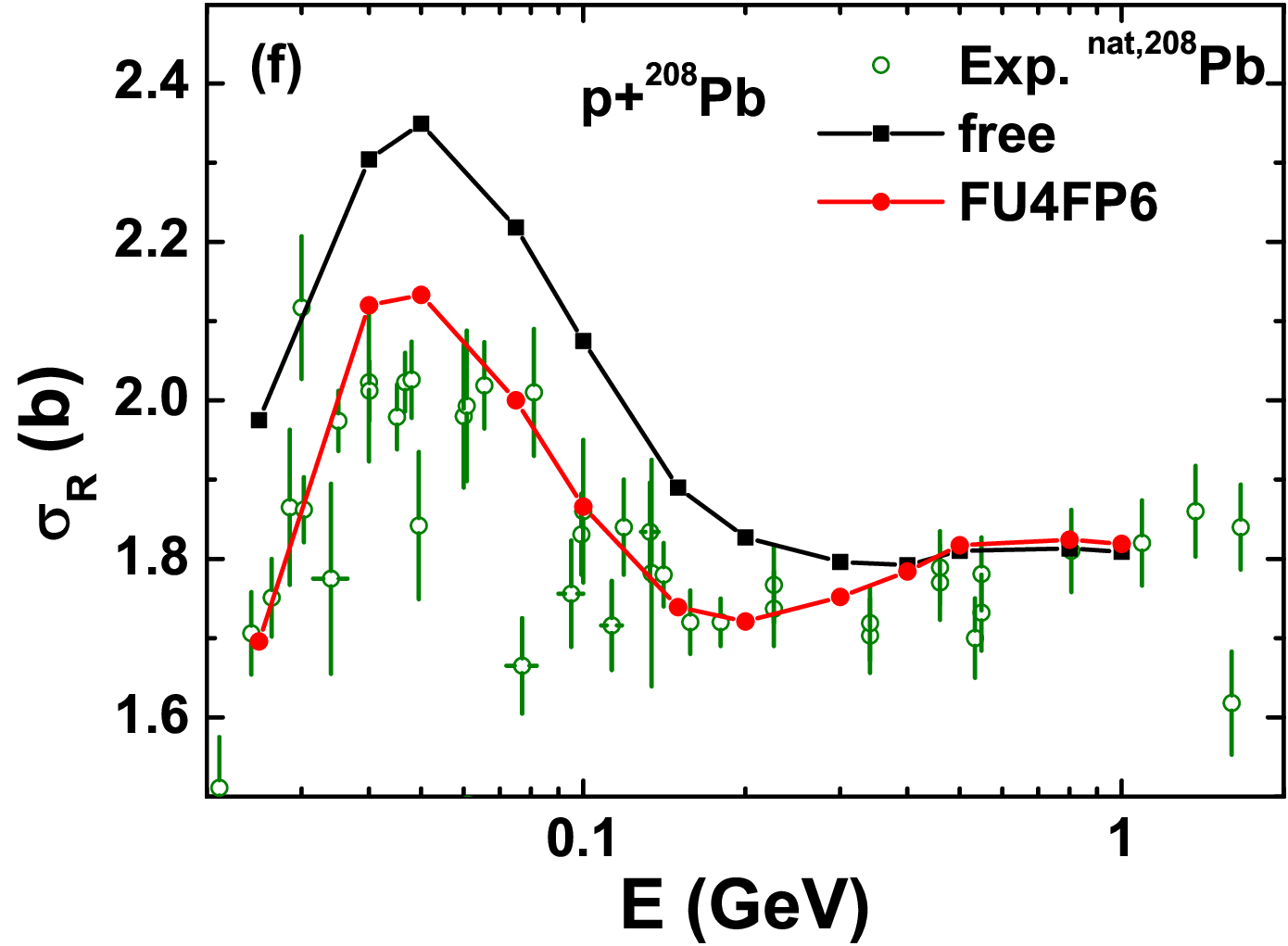}
   \caption{(Color online) Excitation functions of RCS for $p+$A calculated with FU4FP6
   compared with experimental data, respectively. Results calculated with free NNECS are also represented for reference.
   Experimental data are taken form
   Refs.~\cite{Carlson1996,Dietrich2002,Auce2005,Lantz2005,Ingemarsson1999} for C,
 ~\cite{Carlson1996,Dietrich2002} for Ca,~\cite{Carlson1996,Auce2005,Lantz2005,Ingemarsson1999} for Al,
 ~\cite{Carlson1996,Auce2005} for Sn,~\cite{Carlson1996,Ingemarsson1999} for Zr, and
 ~\cite{Carlson1996,Dietrich2002,Auce2005} for Pb targets, respectively.
   }
   \label{pA}
  \end{figure*}

 More experimental data are used to test the obtained in-medium NNECS FU4FP6. From light to heavy targets,
 reactions of proton-induced on $^{12}$C, $^{27}$Al, $^{40,48}$Ca, $^{90}$Zr, $^{118}$Sn, $^{208}$Pb are simulated.
 The excitation functions of RCS for $p+$A calculated with the in-medium NNECS compared with the experimental data
 are presented in Fig.~\ref{pA}.
 One can see that, all experimental data can be quite well reproduced.
 It demonstrates that the in-medium NNECS given by FU4FP6 is reasonable.
 Especially, not only the excitation functions of RCS for the targets along the $\beta$-stable line can be well described,
 the excitation function of RCS for $^{48}$Ca, which is far away from the $\beta$-stable line, can also be well reproduced.
 Beside the proton-induced reactions, the neutron-induced reactions are also used to test the in-medium NNECS.
 \begin{figure*}[hbtp]
  \centering
   \includegraphics[width=0.32\textwidth]{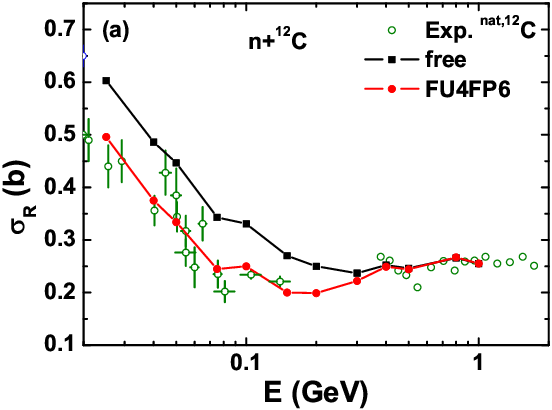}
   \includegraphics[width=0.32\textwidth]{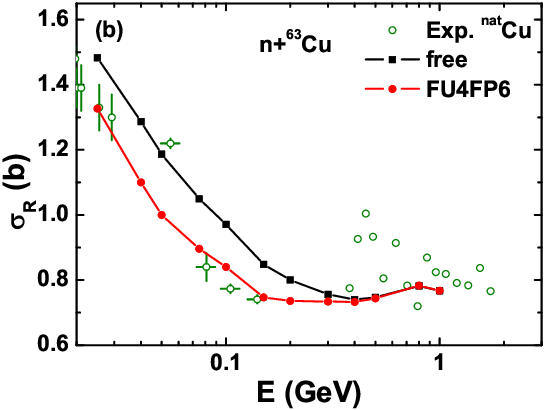}
   \includegraphics[width=0.32\textwidth]{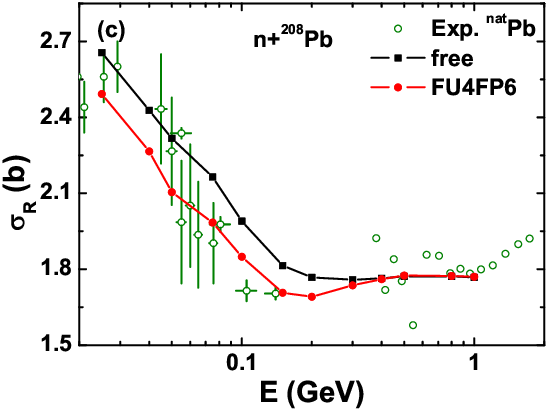}
   \caption{(Color online) Same as Fig.~\ref{pA} but for neutron-induced reactions.
   Experimental data are taken form Refs.
  ~\cite{Zanelli1981,Dimbylow1980,Schimmerling1973,Schimmerling1971,Voss1956,Ibaraki2002} for C,
 ~\cite{Schimmerling1973,Schimmerling1971,MacGregor1958,Voss1956} for Cu, and
 ~\cite{Schimmerling1973,Schimmerling1971,MacGregor1958,Voss1956,Ibaraki2002} for Pb, respectively.}
   \label{nA}
  \end{figure*}
 As the results presented in Fig.~\ref{nA},
 the excitation functions of RCS for $n+$A calculated with FU4FP6 are in good agreement with the experimental data.
 It means that the medium correction on isospin dependence of NNECS is considered reasonably.

\section{Conclusion}

 By using the improved quantum molecular dynamics model,
 the nucleon-induced reactions on various targets with incident energies from 25 MeV to 1 GeV
 are investigated.
 The reaction cross section is found to be very sensitive to the nucleon-nucleon cross sections.
 By comparing the excitation function of reaction cross sections
 between ImQMD model calculations and experimental data,
 the medium modifications on the free nucleon-nucleon elastic cross sections are investigated.
 An isospin, density, and momentum-dependence
 medium correction factor on free nucleon-nucleon elastic cross sections is obtained.
 The parameterized formula is simple to use and gives reliable results for the nuclear reactions
 systems below saturation density.

 \begin{acknowledgments}
 This work has been supported by
 the Natural Science Foundation of Guangxi province under
 Grant No. 2016GXNSFFA380001, 2017GXNSFGA198001,
 the foundation of Guangxi innovative team
 and distinguished scholar in institutions of higher education,
 and
 the National Natural Science Foundation of China under Grant Nos.
 11365004, 
 11475262. 
 \end{acknowledgments}


\begin{thebibliography}{99}
 \bibitem{HanYL1994} Y. Han, G. Mao, Z. Li, and Y. Zhuo, Phys. Rev. C 50, 961 (1994).
 \bibitem{ChouKC1985} K. chao Chou, Z. bin Su, B. lin Hao, and L. Yu, Phys. Rep. 118, 1 (1985).
 \bibitem{Danielewicz1984a} P. Danielewicz, Ann. Phys. 152, 239 (1984).
 \bibitem{Danielewicz1984b} P. Danielewicz, Ann. Phys. 152, 305 (1984).
 \bibitem{Cugnon1987} J. Cugnon, A. Lejeune, and P. Grang\'{e}, Phys. Rev. C 35, 861 (1987).
 \bibitem{Kohler1991} H. K\"{o}hler, Nuclear Physics A 529, 209 (1991).
 \bibitem{Danielewicz2003} L. Shi and P. Danielewicz, Phys. Rev. C 68, 064604 (2003).
 \bibitem{MaoGJ1994} G. Mao, Z. Li, Y. Zhuo, Y. Han, and Z. Yu, Phys. Rev. C 49, 3137 (1994).
 \bibitem{MaoGJ1996} G. Mao, Z. Li, and Y. Zhuo, Phys. Rev. C 53, 2933 (1996).
 \bibitem{LiQF2000} Q. Li, Z. Li, and G. Mao, Phys. Rev. C 62, 014606 (2000).
 \bibitem{LiQF2004} Q. Li, Z. Li, and E. Zhao, Phys. Rev. C 69, 017601 (2004).
 \bibitem{Haar1987} B. ter Haar and R. Malfliet, Phys. Rev. C 36, 1611 (1987).
 \bibitem{Haar1987R} B. ter Haar and R. Malfliet, Phys. Rep. 149, 207 (1987).
 \bibitem{LiGQ1993} G. Q. Li and R. Machleidt, Phys. Rev. C 48, 1702 (1993).
 \bibitem{LiGQ1994} G. Q. Li and R. Machleidt, Phys. Rev. C 49, 566 (1994).
 \bibitem{Fuchs2001} C. Fuchs, A. Faessler, and M. El-Shabshiry, Phys. Rev. C 64, 024003 (2001).
 \bibitem{Kohno1998} M. Kohno, M. Higashi, Y. Watanabe, and M. Kawai, Phys. Rev. C 57, 3495 (1998).
 \bibitem{Alm1994} T. Alm, G. R\"{o}pke, and M. Schmidt, Phys. Rev. C 50, 31 (1994).
 \bibitem{Sammarruca2014} F. Sammarruca, Eur. Phys. J. A (2014).
 \bibitem{LiQF2006} Q. Li, Z. Li, S. Soff, M. Bleicher, and H. St\"ocker, J. Phys. G Nucl. Partic. 32, 407 (2006).
 \bibitem{LiQF2011} Q. Li, C. Shen, C. Guo, Y. Wang, Z. Li, J. Lukasik, and W. Trautmann, Phys. Rev. C 83, 044617 (2011).
 \bibitem{WangYJ2014} Y. Wang, C. Guo, Q. Li, H. Zhang, Z. Li, and W. Trautmann, Phys. Rev. C 89, 034606 (2014).
 \bibitem{Negele1981} J. W. Negele and K. Yazaki, Phys. Rev. Lett. 47, 71 (1981).
 \bibitem{Pandharipande1991} V. R. Pandharipande and S. C. Pieper, Phys. Rev. C 45, 791 (1991).
 \bibitem{Persram2002} D. Persram and C. Gale, Phys. Rev. C 65, 064611 (2002).
 \bibitem{LiBA2005} B.-A. Li and L.-W. Chen, Phys. Rev. C 72, 064611 (2005).
 \bibitem{Westfall1993} G. D. Westfall, W. Bauer, D. Craig, M. Cronqvist, E. Gaultieri, S. Hannuschke, D. Klakow,
      T. Li, T. Reposeur, A. M. Vander Molen, et al., Phys. Rev. Lett. 71, 1986 (1993).
 \bibitem{ZhouHB1993} H. Zhou, Z. Li, and Y. Zhuo, Phys. Lett. B 318, 19 (1993).
 \bibitem{Klakow1993} D. Klakow, G. Welke, and W. Bauer, Phys. Rev. C 48, 1982 (1993).
 \bibitem{ZhangYX2006} Y. Zhang and Z. Li, Phys. Rev. C 74, 014602 (2006).
 \bibitem{ZhangYX2007} Y. Zhang, Z. Li, and P. Danielewicz, Phys. Rev. C 75, 034615 (2007).
 \bibitem{Tanaka1996} E. I. Tanaka, H. Horiuchi, and A. Ono, Phys. Rev. C 54, 3170 (1996).
 \bibitem{Tripathi1998} R. K. Tripathi, F. A. Cucinotta, and J. W. Wilson, Technical Report NASA/TP-1998-208438 (1998).
 \bibitem{Tripathi1999} R. K. Tripathi, F. A. Cucinotta, and J. W. Wilson, Technical Report NASA/TP-1999-209125 (1999).
 \bibitem{Tripathi2001} R. K. Tripathi, J. W. Wilson, and F. A. Cucinotta, Nucl. Instrum. Meth. B 173, 391 (2001).
 \bibitem{OuL2008JPG} L. Ou, Z. Li, and X. Wu, J. Phys. G Nucl. Partic. 35, 055101 (2008).
 \bibitem{OuL2008PRC} L. Ou, Z. Li, and X. Wu, Phys. Rev. C 78, 044609 (2008).
 \bibitem{ChenLW2010} L.-W. Chen, C. M. Ko, B.-A. Li, and J. Xu, Phys. Rev. C 82, 024321 (2010).
 \bibitem{Samyn2002} M. Samyn, S. Goriely, P.-H. Heenen, J. Pearson, and F. Tondeur, Nucl. Phys. A 700, 142 (2002).
 \bibitem{Dobaczewski1984} J. Dobaczewski, H. Flocard, and J. Treiner, Nucl. Phys. A 422, 103 (1984).
 \bibitem{Reinhard1995} P.-G. Reinhard and H. Flocard, Nucl. Phys. A 584, 467 (1995).
 \bibitem{Cugnon1996} J. Cugnon, D. L'H\^{o}te, and J. Vandermeulen, Nucl. Instrum. Meth. B 111, 215 (1996).
 \bibitem{Carlson1996} R. F. Carlson, Atom. Data Nucl. Data 63, 93 (1996).
 \bibitem{Dutra2012} M. Dutra, O. Lourenco, J. S. S\'{a} Martins, A. Delfino, J. R. Stone, and P. D. Stevenson, Phys. Rev. C 85, 035201 (2012).
 \bibitem{Dietrich2002} F. Dietrich, E. Hartouni, S. Johnson, G. Schmid, R. Soltz, W. Abfalterer, R. Haight, L.Waters,
      A. Hanson, R. Finlay, et al., J. Nucl. Sci. Technol. 39 Suppl. 2, 269 (2002).
 \bibitem{Auce2005} A. Auce, A. Ingemarsson, R. Johansson, M. Lantz, G. Tibell, R. F. Carlson, M. J. Shachno,
    A. A. Cowley, G. C. Hillhouse, N. M. Jacobs, et al., Phys. Rev. C 71, 064606 (2005).
 \bibitem{Lantz2005} M. Lantz, M. N. Jacobs, A. Auce, R. F. Carlson, A. A. Cowley, S. V. F\"{o}rtsch, G. C. Hillhouse,
    A. Ingemarsson, R. Johansson, K. J. Lawrie, et al., in International Conference on Nuclear
    Data for Science and Technology, edited by R. C. Haight, M. B. Chadwick, T. Kawano, and
    P. Talou (2005), vol. 769 of American Institute of Physics Conference Series, pp. 846-849.
 \bibitem{Ingemarsson1999} A. Ingemarsson, J. Nyberg, P. Renberg, O. Sundberg, R. Carlson, A. Auce, R. Johansson,
    G. Tibell, B. Clark, L. K. Kerr, et al., Nucl. Phys. A 653, 341 (1999).
 \bibitem{Zanelli1981} C. I. Zanelli, P. P. Urone, J. L. Romero, F. P. Brady, M. L. Johnson, G. A. Needham, J. L.
    Ullmann, and D. L. Johnson, Phys. Rev. C 23, 1015 (1981).
 \bibitem{Dimbylow1980} P. J. Dimbylow, Phys. Med. Biol. 25, 637 (1980).
 \bibitem{Schimmerling1973} W. Schimmerling, T. J. Devlin, W. W. Johnson, K. G. Vosburgh, and R. E. Mischke, Phys.
    Rev. C 7, 248 (1973).
 \bibitem{Schimmerling1971} W. Schimmerling, T. Devlin, W. Johnson, K. Vosburgh, and R. Mischke, Phys. Lett. B 37, 177 (1971).
 \bibitem{Voss1956} R. G. P. Voss and R. Wilson, 236, 41 (1956).
 \bibitem{Ibaraki2002} M. Ibaraki, M. Baba, T. Miura, T. Aqki, T. Hiroishi, H. Nakashima, S. ichiro Meigo, and
        S. Tanaka, J. Nucl. Sci. Technol. 39 Suppl. 2, 405 (2002).
 \bibitem{MacGregor1958} M. H. MacGregor, W. P. Ball, and R. Booth, Phys. Rev. 111, 1155 (1958).
\end{thebibliography}
 \end{document}